\documentclass[twocolumn,floatfix,superscriptaddress,amsmath,amsfonts,amssymb,aps]{revtex4-2}
\usepackage{graphicx}
\usepackage{dcolumn}
\usepackage{bm}
\usepackage{color}
\usepackage{amssymb}
\usepackage{amstext}
\usepackage{amsbsy}
\usepackage{float}

\begin{document}

\preprint{APS/123-QED}

\title{Electronic localization on the structural inhomogeneities formed due to Bi and Te deficiency in the MBE grown films of AFM topological insulator MnBi$_2$Te$_4$:\\  
Evidence from spectroscopic ellipsometry and infrared studies}

\author{N. N. Kovaleva}
\email{kovalevann@lebedev.ru}
\affiliation{P.N. Lebedev Physical Institute, Russian Academy of Sciences, Leninsky prospect 53, 119991 Moscow, Russia}
\author{D. Chvostova}
\affiliation{Institute of Physics, Academy of Sciences of the Czech Republic, Na Slovance 2, 18221 Prague, Czech Republic}
\author{T. N. Fursova}
\affiliation{Institute of Solid State Physics, Russian Academy of Sciences, Academician Osipyan 2, 142432 Chernogolovka, Russia}
\author{A. V. Muratov}
\affiliation{P.N. Lebedev Physical Institute, Russian Academy of Sciences, Leninsky prospect 53, 119991 Moscow, Russia}
\author{S. I. Bozhko}
\affiliation{Institute of Solid State Physics, Russian Academy of Sciences, Academician Osipyan 2, 142432 Chernogolovka, Russia}
\author{\mbox{Yu. A. Aleshchenko}}
\affiliation{P.N. Lebedev Physical Institute, Russian Academy of Sciences, Leninsky prospect 53, 119991 Moscow, Russia}
\author{A. Dejneka}
\affiliation{Institute of Physics, Academy of Sciences of the Czech Republic, Na Slovance 2, 18221 Prague, Czech Republic}
\author{D. V. Ishchenko}
\affiliation{Rzhanov Institute of Semiconductor Physics SB RAS, 630090 Novosibirsk, Russia}
\author{O. E. Tereshchenko}
\affiliation{Rzhanov Institute of Semiconductor Physics SB RAS, 630090 Novosibirsk, Russia}
\affiliation{Department of Physics, Saint Petersburg State University, 198504 St. Petersburg, Russia}
\author{K. I. Kugel}
\affiliation{Institute for Theoretical and Applied Electrodynamics, Russian Academy of Sciences, 125412 Moscow, Russia}
\affiliation{National Research University Higher School of Economics, 101000 Moscow, Russia}
\date{\today}

\begin{abstract}
The intrinsic substitutional and antisite defects cause unintentional doping and shift of the $E_F$ position above the conduction band minimum in the AFM topological insulator (TI) MnBi$_2$Te$_4$. This prevents measurements of the quantum anomalous Hall effect (QAH) and investigation of the topological Dirac states. In the present study, the Mn-Bi-Te films grown by the MBE technique onto Si(111) substrates with decreasing Bi and Te contents and increasing Mn content were investigated by 0.5-6.5\,eV spectroscopic ellipsometry. In addition, the 0.004-0.9\,eV infrared (IR) transmittance spectra were examined. An effective medium model was used to reproduce the measured ellipsometric angles, $\Psi(\omega)$ and $\Delta(\omega)$, of the Mn-Bi-Te films in terms of the constructed model, including film thickness, surface roughness, and volume fractions of two (MnTe and Bi$_2$Te$_3$) or three constituents, the latter being associated with the structural inhomogeneities contribution. The results obtained for the inhomogeneous Mn-Bi-Te films using the three-phase EMA model indicate that the defect-associated optical response systematically shifts to higher photon energies from $\sim$1.95 to $\sim$2.43\,eV with decreasing Te and Bi contents and increasing Mn content, pointing out that the electrons become more deeply localized in the formed structural inhomogeneities. The obtained results indicate that the structure of the non-stoichiometric Mn-Bi-Te films is not continuous but represented by regions of nearly stoichiometric MnBi$_2$Te$_4$ phase, which includes hollows or quantum anti-dots (QADs). The measured FIR transmittance spectra for the non-stoichiometric Mn-Bi-Te films show substantially reduced (or absent) contribution(s) from free charge carriers, which supports the relevance of localization effects.
\end{abstract}

\pacs{Valid PACS appear here}

\maketitle

\section{\label{sec:level1}INTRODUCTION}
Intrinsic antiferromagnetic (AFM) topological insulators (TIs) belonging to a homological series MnTe$\cdot n$Bi$_2$Te$_3$ ($n$\,=\,1,2,3, ... ) are promising materials for accomplishing curious topological quantum states \cite{Otrokov,Rusinov,Zhang,Aliev,Jahangirli,Frolov}. Additionally, advanced magnetic and topological properties can be engineered in [(MnTe$\cdot$Bi$_2$Te$_3$)(MnTe)$_m$]$_N$ superlattices grown by the MBE technique \cite{Chen}, as well as in superlattices represented by sub- and nanosize periodic arrays, including layers of ferromagnetic (FM) and TI materials \cite{Kovaleva_APL_2021,Kovaleva_Coatings_2022}. 
In general, TIs possess bulk gaps and metallic Dirac surface states having linear dispersion, which are protected by time reversal symmetry (TRS). In the  primary compound MnBi$_2$Te$_4$, Mn ions have the intralayer FM coupling, while the interlayer interaction gives rise to the A-type AFM coupling. The intrinsic magnetic ordering can break the TRS and open an exchange gap leading to the quantum anomalous Hall (QAH) effect \cite{Yu,CZ_Chang}. The key condition for experimental observation of the QAH effect is the location of the Fermi level ($E_F$) within the bulk band gap. Yet, high concentrations of intrinsic defects in MnBi$_2$Te$_4$ cause a pronounced metallic conductivity affecting the $E_F$ position, magnetism, and topological properties. The topological Dirac-cone states have reliably been resolved in MnBi$_2$Te$_4$ by angle-resolved photoemission spectroscopy (ARPES) revealing that the $E_F$ is located above the conduction band bottom \cite{Otrokov,Rusinov,Hao,Swatek,Vidal,Glazkova}. 

As a well-known TI Bi$_2$Te$_3$, MnBi$_2$Te$_4$ crystallizes into a rhombohedral tetradimite structure (space group $R{\bar 3}m$) with unit cell parameters $a$\,=\,$b$\,=\,4.3304(4)\,$\AA$, $c$\,=\,40.919(4)\,$\AA$, $\alpha$\,=\,$\beta$\,=\,90$^\circ$, $\gamma$\,=\,120$^\circ$ \cite{Lee,Aliev}. Along the $c$-axis direction, MnBi$_2$Te$_4$ exhibits the -7-7-7- sequence of the Te-Bi-Te-Mn-Te-Bi-Te septuple-layer (SL) blocks. In each SL block, the hexagonal MnTe layer is intercalated within the hexagonal Bi$_2$Te$_3$ layer to form a natural internal heterostructure, being under significant tensile strain in the MnBi$_2$Te$_4$ crystal structure. The strain in the internal heterostructure stimulates the generation of high density donor-type substitutional Bi$^{\circ}_{\rm Mn}$ defects, which are responsible for $n$-type conductivity. Since Mn and Bi atoms occupy the cation positions in the disordered manner forming mixed atomic layers, a high level of disorder can be characteristic of the MnBi$_2$Te$_4$ structure. In addition to the disorder in the cation sublattices, an anti-structural (antisite) disorder can exist, when the Te atoms partially occupy the Bi atom positions, and vice versa. Thus, in the MnBi$_2$Te$_4$ structure, the substitutional defects Bi$^{\circ}_{\rm Mn}$ and Mn$^{\prime}_{\rm Bi}$, of donor and acceptor types, respectively, as well as the antisite defects Bi$^{\prime}_{\rm Te}$ and Te$^{\circ}_{\rm Bi}$ can exist. The electroneutrality is satisfied when 
\begin{equation}
n+[{\rm Bi}^{\prime}_{\rm Te}]+[{\rm Mn}^{\prime}_{\rm Bi}]+2[{\rm Mn}^{\prime \prime}_{\rm Te}] \rightleftharpoons [{\rm Bi}^{\circ}_{\rm Mn}]+[{\rm Te}^{\circ}_{\rm Bi}]+p, 
\label{eq1}
\end{equation}
where $n$ and $p$ are the concentrations of electrons and holes, respectively. Concentrations of the antisite defects of donor and acceptor types are approximately equal, and then the $n$- or $p$-type of the conductivity is associated with the dominating type of substitutional defects of the donor Bi$^{\circ}_{\rm Mn}$ or acceptor Mn$^{\prime}_{\rm Bi}$ types. 

Moreover, as it was reported earlier, the conductivity of the similar triple compound GeBi$_2$Te$_4$ essentially depends on the deviations $\delta_1$, $\delta_2$, and $\delta_3$ from the stoichiometry in Ge$_{1\pm\delta_1}$Bi$_{2+\delta_2}$Te$_{4+\delta_3}$ \cite{Shelimova}. Thus, the change of $\delta_1$ can lead to the inversion of the conductivity sign from the $p$-type for the deficiency of Ge to $n$-type for the excess of Ge. The conductivity is low for the alloy composition Ge$_{1.2}$Bi$_2$Te$_4$, which is proposed to be arranged in the nine-layer (NL) crystal structure with $c$\,=\,17.3 $\AA$ \cite{Karpinsky_1997}. However, the homogeneity range in the system Ge-Bi-Te is rather narrow for Ge$_{1\pm\delta_1}$Bi$_{2+\delta_2}$Te$_{4+\delta_3}$, where the solubility for Ge is anomalously high of $-0.04\,\leq\,\delta_1\,\leq\,0.30$, while the solubilities for Bi and Te are notably lower, being positive of $0\,\leq\,\delta_2\,\leq\,0.05$ and $0\,\leq\,\delta_3\,\leq\,0.14$, respectively \cite{Shelimova}.

Similarly, the homogeneity range is quite narrow in MnBi$_2$Te$_4$ alloy. One can guess that strong deficiency of the Bi and Te constituents in the nonstoichiometric Mn-Bi-Te alloy will result in the formation of structural inhomogeneities in the form of holes leading to the continuity collapse within the Bi and Te layered structure. Then, a puzzle of such structural inhomogeneities will be formed depending on the given Bi and Te deficiency. The morphology profile corresponding to the formed hole pattern will naturally be reproduced along the $c$ axis within the SL structure and, possibly, even in the upper subsequently grown SL blocks. The formed inhomogeneous potential of such structural defects will necessarily lead to the localization of itinerant charge carriers originating from intrinsic doping. Actually, this can help to maintain the $E_F$ position within the bulk band gap in the rest of MnBi$_2$Te$_4$ matrix.             

In the present study, using 0.5-6.5\,eV spectroscopic ellipsometry (SE) we investigate the effective complex dielectric function, represented by the real, 
$\varepsilon_1(\omega)$, and imaginary, $\varepsilon_2(\omega)$, parts for the Mn-Bi-Te films grown by the MBE technique with decreasing Bi and Te contents and increasing Mn content. The quantitative microstructural information of the grown heterogeneous Mn-Bi-Te films can be deduced using the Bruggeman EMA approach \cite{Bruggeman} when accurate dielectric function spectra are available in the visible-near-UV photon energy range for the composite Mn-Bi-Te film sample and those of the two constituents -- MnTe and Bi$_2$Te$_3$. An effective medium model was used to reproduce the complex dielectric function spectra of the composite in terms of the constructed model, including film thickness, surface roughness, and volume fractions of two or three constituents, which are determined from the best least-squares fit of the calculated spectra to the measured ones \cite{Aspnes}.

We found that while the maximum values of $\varepsilon_1(\omega)$ and $\varepsilon_2(\omega)$ decrease with decreasing Bi and Te contents and increasing Mn content in the nonstoichiometric Mn-Bi-Te films, the $\varepsilon_2(\omega)$ maximum progressively shifts to higher photon energies from $\sim$1.2 to $\sim$3.7\,eV. The discovered effect is attributed to the third-phase component, which can be associated with the contribution of structural inhomogeneities. In addition, the infrared (IR) transmittance spectra were measured for the grown Mn-Bi-Te film samples. The decreased transmittance for the stoichiometric MnBi$_2$Te$_4$ film sample is observed in the far-IR (FIR) at $\omega$\,$\rightarrow$\,0, which can be related to the emergent Drude-type contribution(s). However, the FIR transmittance increases for the heterogeneous Mn-Bi-Te films suggesting the suppressed Drude-type contribuion(s) in accordance with Eq.\,(\ref{eq1}). The nanocrystalline quality of the investigated MBE grown Mn-Bi-Te films is demonstrated by measurements of the IR transmittance spectra, where the IR phonon modes characteristic of MnTe and MnBi$_2$Te$_4$ crystalline structures were identified. The present study paves the way for optical control of the structural inhomogeneities contribution recognized by the energy levels of localized electrons, as well as of the intrinsic doping in the MBE grown Mn-Bi-Te films and in the MnTe-intercalated [(MnTe$\cdot$Bi$_2$Te$_3$)(MnTe)$_m$]$_N$ superlattices.

\begin{figure}
\includegraphics [width=1.0\columnwidth]{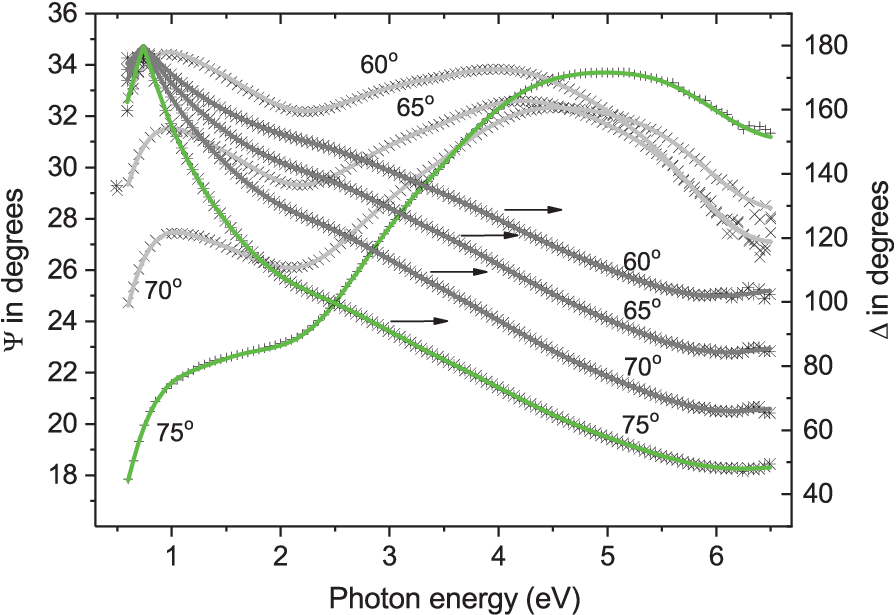}
\caption{Spectra of ellipsometric angles, $\Psi(\omega)$ and $\Delta(\omega)$, measured for the MnBi$_2$Te$_4$/Si(111) film sample (1) at different angles of an incident light of 60$^\circ$, 65$^\circ$, 70$^\circ$, and 75$^\circ$ (shown by symbols). The results of the two-layer model simulations (see the text for details) are displayed by solid curves, where the solid green curve depicts the fitting result for 75$^\circ$.
\vspace{-0.5cm}}
\label{PsiDelta_1}
\end{figure}

\begin{figure*} \centering
\includegraphics[width=16.0cm]{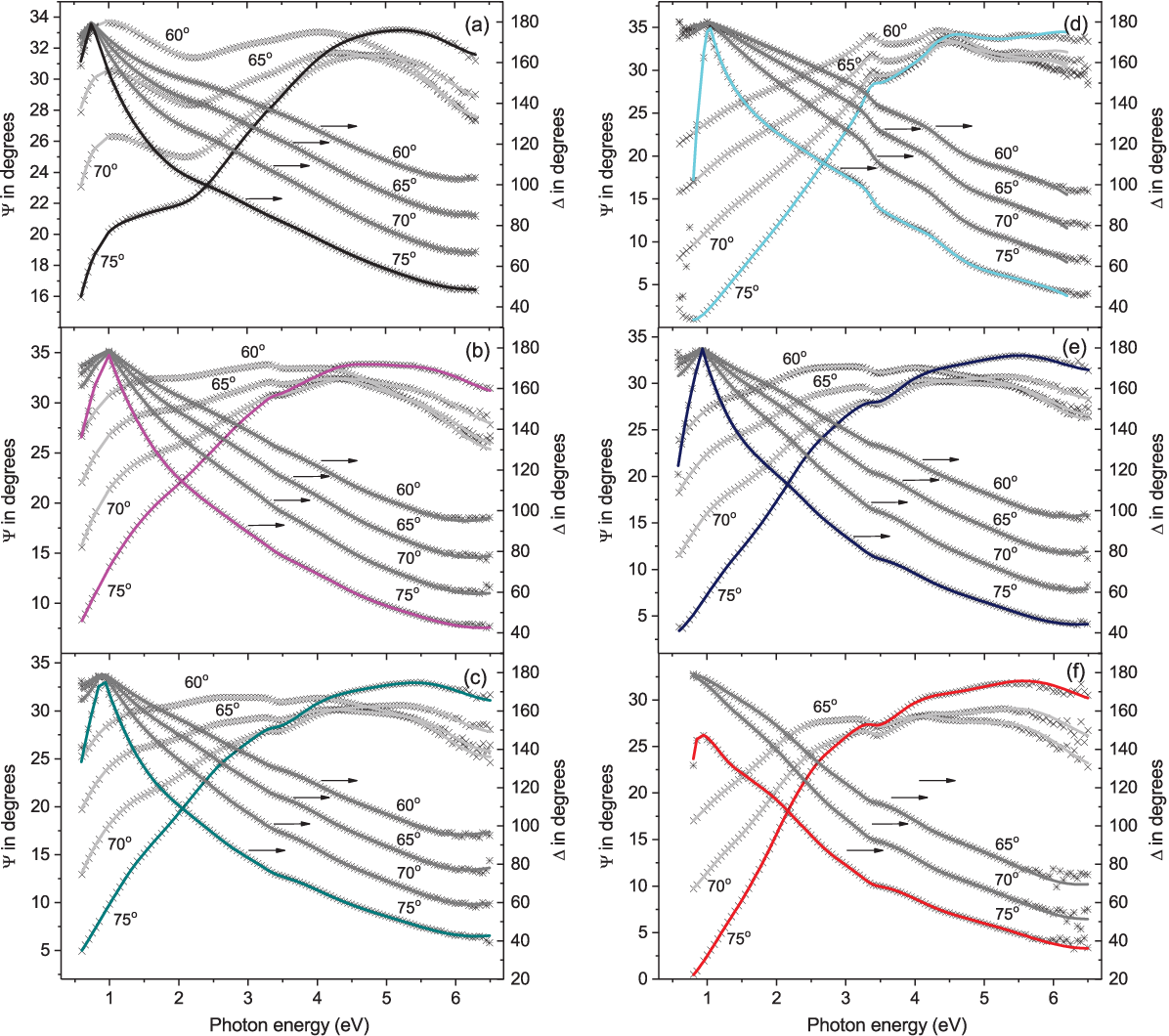}
\caption{Spectra of ellipsometric angles, $\Psi(\omega)$ and $\Delta(\omega)$, measured for the Mn-Bi-Te film samples (a) MnBi$_{1.85}$Te$_{3.8}$/Si (2), (b) MnBi$_{1.6}$Te$_{2.9}$/Si (3), (c) Mn$_2$BiTe$_2$/Si (4), (d) Mn$_{1.3}$BiTe$_{2.5}$/Si (5), (e) Mn$_2$BiTe$_4$/Si (6), and (f) Mn$_2$Bi$_{1.2}$Te$_{4.2}$/Si (7) at different angles of an incident light of 60$^\circ$, 65$^\circ$, 70$^\circ$, and 75$^\circ$ (shown by symbols). The results of the two- or three-layer model simulations are displayed by solid curves, where the solid color curves depict the fitting result for 75$^\circ$.
\vspace{-0.5cm}}
\label{PsiDelta_a_f}
\end{figure*}

\section{\label{sec:level1}SPECTROSCOPIC ELLIPSOMETRY STUDY OF Mn-Bi-Te FILMS}
\subsection{\label{sec:level2_1}MBE growth of Mn-Bi-Te film samples}
The Mn-Bi-Te film samples were prepared utilizing an Angara-type MBE equipment \cite{Tereshchenko}. For film deposition, the sources of Mn (4N 99.99\%), Te (6N 99.9999\%), and the Bi$_2$Te$_3$ compound preliminary synthesized by the Bridgman technique were used. The Mn-Bi-Te films were grown onto the Si(111) substrates, which were RCA cleaned before. The Si(111) substrates were purged of a $\sim$2.5\,nm thick SiO$_2$ surface layer at 750$^\circ$C \textit{in situ} under the ultra-high vacuum conditions in the MBE chamber. The procedure was monitored by the appearance of the Kikuchi lines accompanying the Si(111) 7\,$\times$\,7 structure reconstruction. Then, the Si(111) substrates were got cool to 320\,$^\circ$C, where the dangling Si bonds were passivated in Te flux for approximately 30 s. While the shutters over the Mn, Te, and Bi$_2$Te$_3$ sources were open, the growth of the Mn-Bi-Te films started. The growth rate was monitored using a quartz crystal microbalance (QCM). The crystal structure of the prepared Mn-Bi-Te films was analyzed \textit{in situ} using reflective high energy electron diffraction (RHEED). The elemental composition was verified by X-ray photoelectron spectroscopy (XPS) (for more detail, see Refs.\,\cite{Tereshchenko,Stepina}). For current investigations, the Mn-Bi-Te films -- (1) MnBi$_2$Te$_4$/Si(111), (2) MnBi$_{1.85}$Te$_{3.8}$/Si(111), (3) MnBi$_{1.6}$Te$_{2.9}$/Si(111), (4) Mn$_2$BiTe$_2$/Si(111), (5) Mn$_{1.3}$BiTe$_{2.5}$/Si(111), (6) Mn$_2$BiTe$_4$/Si(111), and (7) Mn$_2$Bi$_{1.2}$Te$_{4.2}$/Si(111) -- with decreasing Te and Bi contents and increasing Mn content were grown. The thicknesses $d_{\rm AFM}$ of the grown films were evaluated from the AFM measurements of a scratch profile at the film edge (for details, see Table\,\ref{tab:table1}).

\subsection{\label{sec:level2_2}Spectroscopic ellipsometry approach for the MBE grown Mn-Bi-Te film samples}
In the course of SE measurements of the Mn-Bi-Te film samples grown by the MBE technique, the spectral dependences of ellipsometric angles, ${\rm \Psi}(\omega)$ and ${\rm \Delta}(\omega)$, were gained at room temperature at several angles of light incidence of 60$^\circ$, 65$^\circ$, 70$^\circ$, and 75$^\circ$ in the 0.5-6.5\,eV spectral range (with the 0.05\,eV spectral resolution) utilizing a J.A. Woollam VASE apparatus. The collected spectra of ${\rm \Psi}(\omega)$ and ${\rm \Delta}(\omega)$ for the investigated series of the Mn-Bi-Te film samples (1) and (2)\,-\,(7) are displayed in Figs.\,\ref{PsiDelta_1} and \ref{PsiDelta_a_f}(a-f), respectively. The ellipsometric angles, ${\rm \Psi}(\omega)$ and ${\rm \Delta}(\omega)$, were simulated using a two-layer model representing the Mn-Bi-Te film on the Si(111) substrate or a three-layer model, which allowed one to include feasible film surface roughness (sr) applying the VASE software \cite{WVASE_software}. For the present model simulations, the Si substrate optical constants ``si$_{-}$jaw2'' available in the VASE database \cite{WVASE_software} were used. The feasible sr layer was considered in the Bruggeman effective medium approximation (EMA) (50\% Mn-Bi-Te film -- 50\% voids). 

Complex dielectric function, $\varepsilon_1(\omega)$+i$\varepsilon_2(\omega)$, of the investigated Mn-Bi-Te films was treated within the generalized oscillator layer \cite{WVASE_software} incorporating the high frequency dielectric constant $\varepsilon_{\infty}$ and a set of Gaussian oscillators, represented by the Kramers-Kronig (KK) consistent parts in $\varepsilon_2$ and $\varepsilon_1$, respectively   
\begin{eqnarray}
\varepsilon_2=Ae^{-(\frac{E-E_0}{\sigma})^2}-Ae^{-(\frac{E+E_0}{\sigma})^2},
\label{DispAnae2}
\end{eqnarray}
\begin{eqnarray}
\varepsilon_1=
\frac{2}{\pi}P\int_0^\infty \frac{\xi\varepsilon_2(\xi)}{\xi^2-E^2}d\xi.
\label{DispAnae1}
\end{eqnarray}
Here, each Gaussian oscillator is characterized by the energy maximum $E_0$, dimensionless oscillator strength $A$, and $\sigma=\frac{Br_0}{2\sqrt{ln(2)}}$, where the $Br_0$ corresponds to the band full width at half maximum (FWHM). In the simulations, the fitting parameters for a $j$th Gaussian oscillator were $E_j$, $Br_j$, and $A_j$.

\begin{figure} 
\includegraphics [width=0.9\columnwidth]{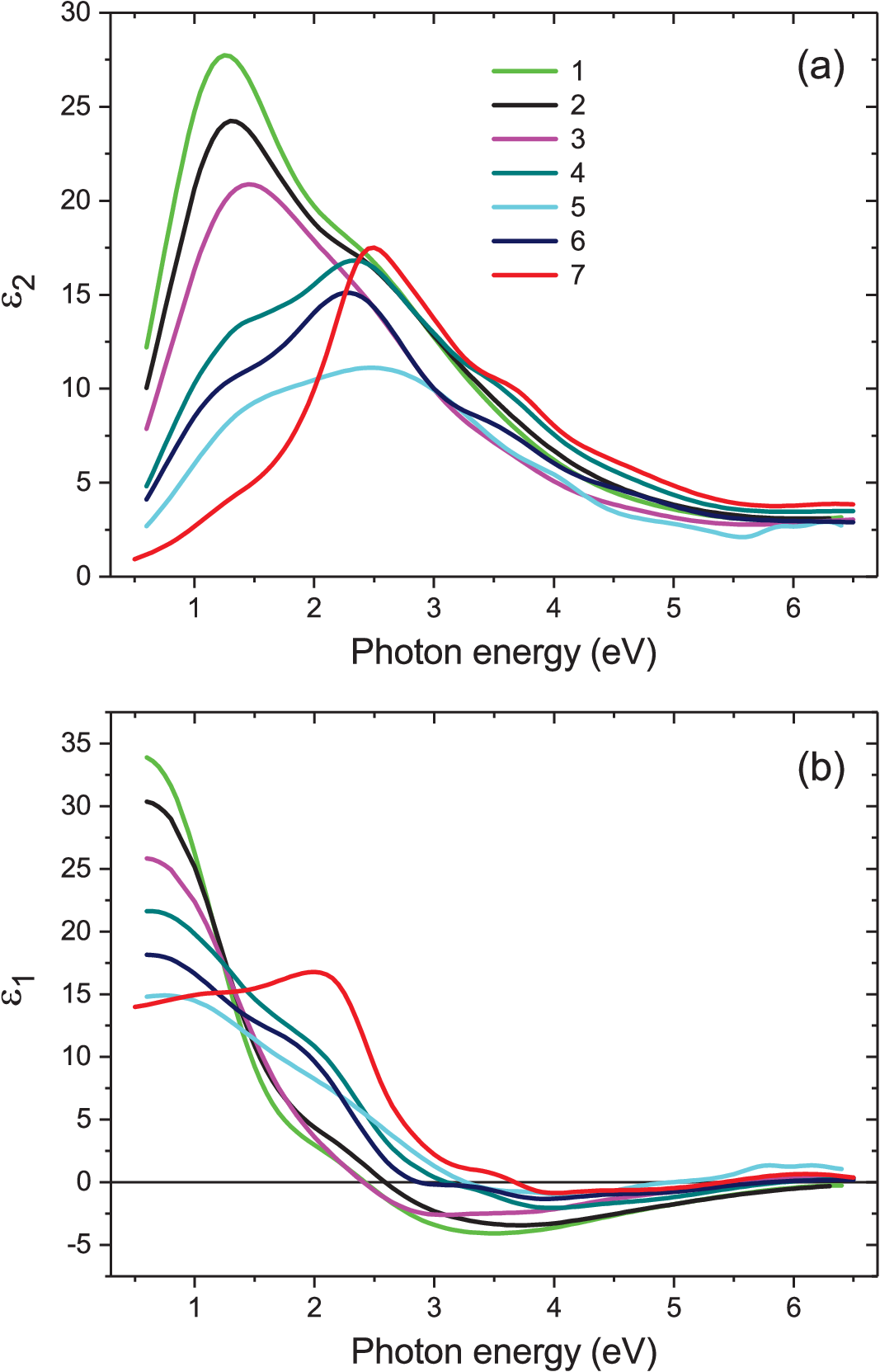}
\caption{(a) Imaginary and (b) real parts of the effective complex dielectric function determined from the model simulations for the studied film series 
(1) MnBi$_2$Te$_4$, (2) MnBi$_{1.85}$Te$_{3.8}$, (3) MnBi$_{1.6}$Te$_{2.9}$, (4) Mn$_2$BiTe$_2$, (5) Mn$_{1.3}$BiTe$_{2.5}$, (6) Mn$_2$BiTe$_4$, and (7) Mn$_2$Bi$_{1.2}$Te$_{4.2}$.
\vspace{-0.5cm}}
\label{e1e2_all}
\end{figure}

\begin{figure*} \centering
\includegraphics[width=13.5cm]{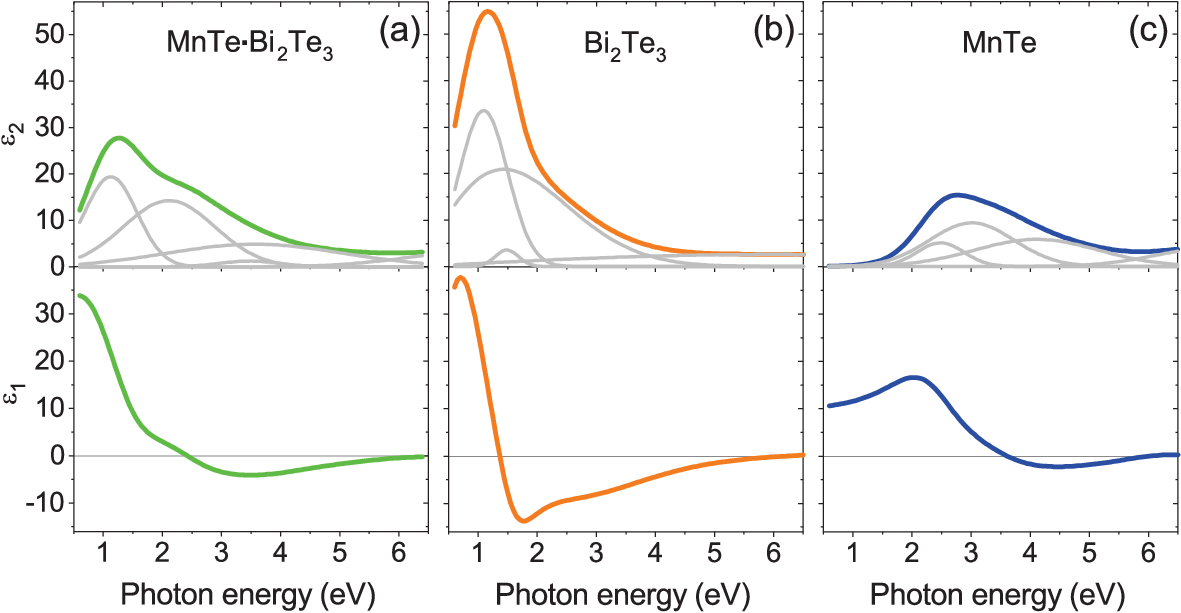}
\caption{The imaginary $\varepsilon_2(\omega)$ (top panels) and real 
$\varepsilon_1(\omega)$ (bottom panels) parts of the complex dielectric function for (a) the stoichiometric MnBi$_2$Te$_4$ film (1), (b,c) the Bi$_2$Te$_3$ and MnTe phases, consistent with the applied EMA model. The contributions from the Gaussian optical bands (see Table\,\ref{tab:table1}) are shown by solid gray curves.    
\vspace{-0.5cm}}
\label{EMA_N164}
\end{figure*}

Using the same approach as in our recent SE study of the MBE grown Mn-Bi-Te film samples MnTe$\cdot$Bi$_{0.12}$Te$_{0.18}$/Si(111), MnTe$\cdot$Bi$_{0.7}$Te/Si(111), and MnTe$\cdot$Bi$_2$Te$_3$/Si(111) \cite{Kovaleva_APL_2024}, we demonstrated that the complex index of refraction of the investigated Mn-Bi-Te films, $n(\omega)$ and $k(\omega)$, as well as the complex dielectric function, $\varepsilon_1(\omega)$+i$\varepsilon_2(\omega)$, can successfully be described in the framework of the two- or three-layer model. The used simulation model suggested that the thicknesses of the Mn-Bi-Te film and sr layer varied until the minimum of the mean-least square error was reached. If the resulting sr layer thickness was equal to 0, the layer was excluded from the model (for more detail, see Refs.\,\cite{Kovaleva_APL_2015,Kovaleva_APL_2017,Kovaleva_Metals_2017}).      
In the present study, which is the follow-up of our recent study \cite{Kovaleva_APL_2024}, we extend the series of Mn-Bi-Te film samples with new samples (2)-(7) synthesized by the MBE technique under the variation of the Bi, Te, and Mn contents. From Fig.\,\ref{PsiDelta_a_f}(a-f) one can see that the measured ellipsometric angles are well described by the model simulations. The resulting Mn-Bi-Te film thicknesses, $d_{\rm eff}$, as well as the sr layer thicknesses (where appropriate), are listed in Table\,\ref{tab:table1}. One can notice that the film thicknesses attested by the AFM measurements, $d_{\rm AFM}$, are reasonably well reproduced by the present model simulations. 

The spectra of the effective complex dielectric function represented by the imaginary $\varepsilon_2(\omega)$ and real $\varepsilon_1(\omega)$ parts, were obtained, which are shown in Fig.\,\ref{e1e2_all}(a,b), respectively. One can see an obvious trend in the $\varepsilon_2(\omega)$ and $\varepsilon_1(\omega)$ spectra with the variation of the Bi, Te, and Mn contents in the investigated Mn-Bi-Te film series (1)-(7). That is, once the values of $\varepsilon_1(\omega)$ and $\varepsilon_2(\omega)$ decrease for the Mn-Bi-Te film series from (1) to (7), the maximum of the $\varepsilon_2(\omega)$ spectra switches from $\sim$1.2-1.5\,eV in the films (1)-(3) to $\sim$2.3-2.6\,eV in the films (4), (6), and (7). Interestingly, for the Mn$_{1.3}$Bi$_2$Te$_{2.5}$ (5), both maxima seem to be nearly equally pronounced. For the stoichiometric MnTe$\cdot$Bi$_2$Te$_3$ film (1), the $\varepsilon_1(\omega)$ varies between 35 at low photon energies and --5 near 3.5\,eV, and the $\varepsilon_2(\omega)$ exhibits the maximum of about 27 near 1.2\,eV, being in good agreement with the SE results for MnBi$_2$Te$_4$ single crystal \cite{Jahangirli}. According to our previous SE study of the Mn-Bi-Te films \cite{Kovaleva_APL_2024}, the dielectric function spectra of the stoichiometric MnTe$\cdot$Bi$_2$Te$_3$ film (1) includes the pronounced interband optical transitions at $\sim$1.1 and $\sim$2.1\,eV. Yet, in the dielectric function spectra of the MnTe$\cdot$Bi$_{0.7}$Te film, the low-energy band at $\sim$1.1\,eV is most completely suppressed and the main interband optical transition appears at $\sim$2.6\,eV. Meanwhile, the higher-energy band peaking at $\sim$3.5-3.7\,eV, which is the predominant optical band in the dielectric function response of the MnTe$\cdot$Bi$_{0.12}$Te$_{0.18}$ film, appears in the investigated Mn-Bi-Te film series for the samples (6) and (7), seemingly being responsible for the MnTe contribution (for more detail, as well as for the Gaussian band parameters resulting from the model simulations, see Table I in Ref.\,\cite{Kovaleva_APL_2024}).              

\subsection{\label{sec:level1}Effective medium approximation for the MBE grown Mn-Bi-Te films}
Since the homogeneity range is rather narrow with respect to the stoichiometry deviations in the MnBi$_2$Te$_4$ compound, strong deficiency of the Bi and/or Te contents in the grown Mn-Bi-Te films possibly leads to substantial corruption in the structure of the Bi and/or Te layers resulting in the formation of structural inhomogeneities in the form of holes or even phase separation. Meantime, an excess of Mn content in the grown Mn-Bi-Te films can lead to the intercalation of MnTe layers inside the film structure. Here, we attempt to describe the investigated inhomogeneous Mn-Bi-Te films representing their macroscopic dielectric response $\varepsilon$ associated with a two- or three-phase composite in the framework of the Bruggeman EMA approach. Indeed, an EMA can be used to reproduce the complex dielectric function spectra of the composite in terms of the constructed model, including film thickness, surface roughness, and volume fractions of the constituents, which are determined from the best least-squares fit of the simulated spectra to the measured ones \cite{Aspnes}.

\begin{table}
\caption{\label{tab:table1}
Results from the model simulation (see the text for details) for the investigated series of the Mn-Bi-Te film samples.}
\begin{ruledtabular}
\begin{tabular}{llllccc}
Mn-Bi-Te film            & $d_{\rm AFM}$ & $d_{\rm eff}$ & $d_{\rm EMA}$ & Bi$_2$Te$_3$ & MnTe      \\
                          & (nm) & (nm)              &(nm)            & \% &  \%   \\ \hline
(1) MnBi$_2$Te$_4$        &  45 & 44.6\,$\pm$\,0.2   & 44.4\,$\pm$\,0.1    & 60.1 & 39.9  \\
(2) MnBi$_{1.85}$Te$_{3.8}$&  40 & 46.3\,$\pm$\,0.6   & 45.64\,$\pm$\,0.1  & 53.1 & 39.0  \\
(3) MnBi$_{1.6}$Te$_{2.9}$&  20 & 27.1\,$\pm$\,1.4   & 29.7\,$\pm$\,0.5    & 41.5 & 45.2  \\
                          &     &                    & sr\,1.0\,$\pm$\,0.3 &      &       \\
(4) Mn$_2$BiTe$_2$        &  25 & 27.1\,$\pm$\,0.5   & 41.3\,$\pm$\,1.0    & 27.6 & 46.2  \\
                          &     & sr\,4.4\,$\pm$\,0.1&                     &      & \\
(5) Mn$_{1.3}$BiTe$_{2.5}$  &  20 & 14.8\,$\pm$\,0.1   & 14.9\,$\pm$\,0.1    & 21.4 & 41.3  \\
(6) Mn$_2$BiTe$_4$        &  25 & 27.7\,$\pm$\,0.6   & 34.7\,$\pm$\,0.5    & 15.8 & 50.0  \\
(7) Mn$_2$Bi$_{1.2}$Te$_{4.2}$    &  25 & 27.6\,$\pm$\,0.9   & 28.6\,$\pm$\,0.6    & 5.8  & 54.4  \\
                          &     & sr\,6.7\,$\pm$\,0.3& sr\,6.8\,$\pm$\,0.4 &      &       \\
\end{tabular}
\end{ruledtabular}
\end{table}

\subsubsection{\label{sec:level2_1}Two-phase EMA model for MnTe$\cdot$Bi$_2$Te$_3$}
Initially, we design the KK-consistent model for the complex dielectric function of the Bi$_2$Te$_3$ phase, according to the spectra of Bi$_2$Te$_3$ single crystal presented in Fig.\,2(a,b) in Ref.\,\cite{Jahangirli}. Next, using the two-phase Bruggeman EMA model for the generalized oscillator layer, which describes the stoichiometric MnTe$\cdot$Bi$_2$Te$_3$ film (1) and incorporating the KK consistent complex dielectric function of the Bi$_2$Te$_3$ phase, we determine the complex dielectric function of the MnTe phase within the stoichiometric MnTe$\cdot$Bi$_2$Te$_3$ film (1). The fit quality in the framework of the two-phase EMA model is in good agreement with the measured ellipsometric angles (see Fig.\,\ref{PsiDelta_1}), as well as with the complex dielectric function of the stoichiometric MnBi$_2$Te$_4$ film (1) presented in Fig.\,\ref{e1e2_all}(a,b). The results of the best-fit EMA simulations determine the volume fractions of the MnTe and Bi$_2$Te$_3$ phases in the stoichiometric MnTe$\cdot$Bi$_2$Te$_3$ film (1): 40\% MnTe\,+\,60\% Bi$_2$Te$_3$. The Gaussian oscillator parameters, $E_j$, $Br_j$, and $A_j$, derived from the EMA simulations are given in Table\,\ref{tab:table2}, and the imaginary $\varepsilon_2(\omega)$ and real $\varepsilon_1(\omega)$ parts of the complex dielectric function for the MnBi$_2$Te$_4$, Bi$_2$Te$_3$, and MnTe phases consistent with the present EMA simulations are shown in Fig.\,\ref{EMA_N164} (a)-(c), respectively. 

The elaborated dispersion model for the MnTe includes three Gaussian bands at 2.5, 3.0, and 4.1\,eV, where the most strong interband optical transition is predicted at 3.0\,eV. The presence of these optical bands in the dielectric function of the investigated Mn-Bi-Te films is seemingly responsible for the MnTe contribution. We obtain that for the MnTe the $\varepsilon_1(\omega)$ varies between 10 at low photon energies and --2.2 near 4.5\,eV, and the $\varepsilon_2(\omega)$ exhibits the maximum about 15.4 near 2.8\,eV (see Fig.\,\ref{EMA_N164}(c)). We would like to note that the obtained complex dielectric function of the MnTe is in good agreement with the theoretical values of $\varepsilon_1^{xx}(\omega)$ and $\varepsilon_2^{xx}(\omega)$ calculated for MnTe single crystal (see Fig.\,5(a,b) of Ref.\,\cite{Mazin}).

\begin{figure}
\includegraphics [width=0.9\columnwidth]{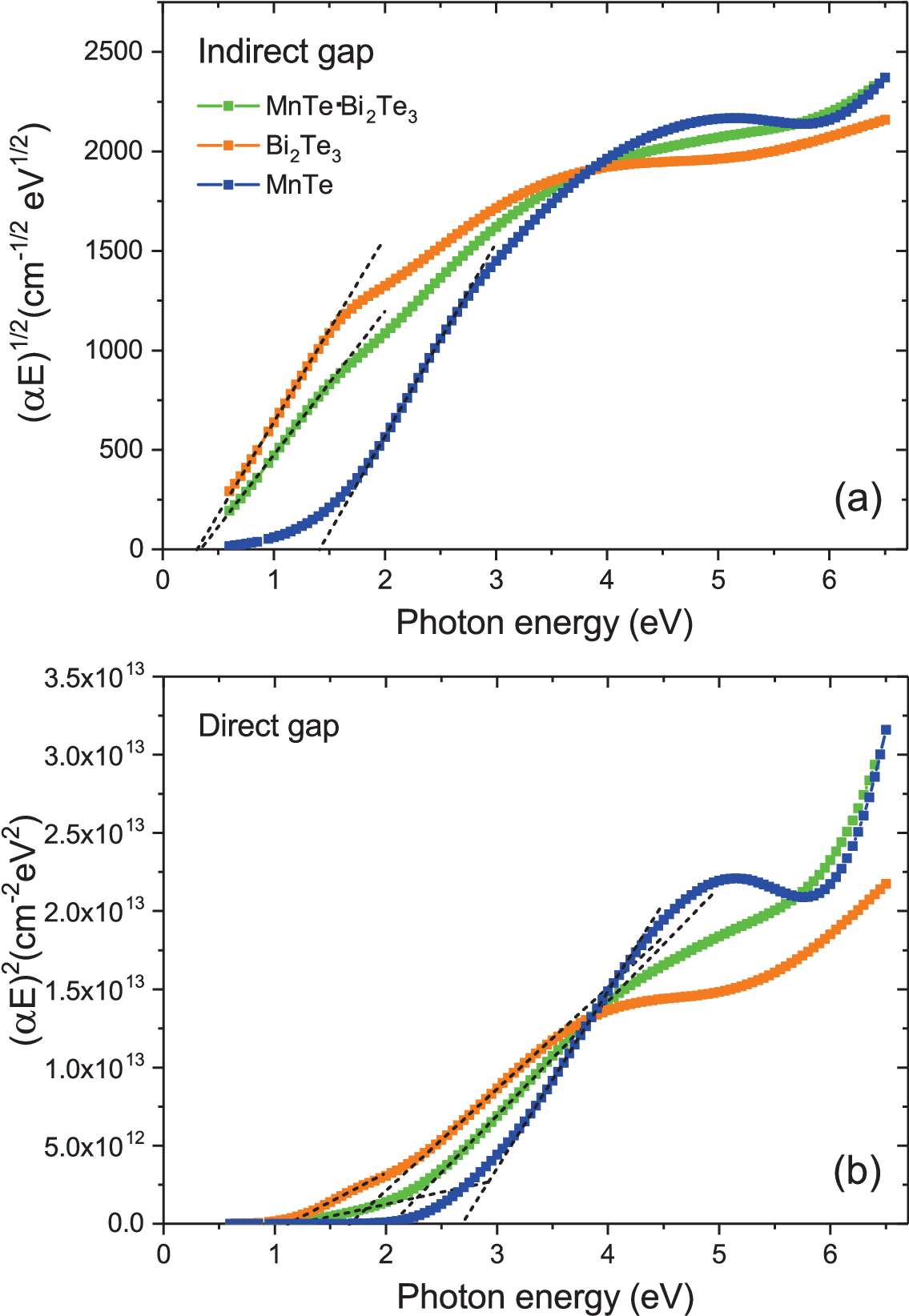}
\caption{(a) Indirect and (b) direct gap values estimated for the stoichiometric MnBi$_2$Te$_4$ film (1), the Bi$_2$Te$_3$ single crystal, and the MnTe constituent (for more detail, see the text, Fig.\,\ref{EMA_N164}, and Table\,\ref{tab:table2}).
\vspace{-0.5cm}}
\label{Gaps}
\end{figure}

From the obtained EMA results, the absorption coefficient $\alpha(\omega)=4 \pi k(\omega)\omega$ was calculated from the respective extinction coefficient $k(\omega)$ for the MnBi$_2$Te$_4$, Bi$_2$Te$_3$, and MnTe phases. Next, the indirect gap values were estimated using the linear extrapolation of the $[\alpha(\omega)\omega]^{1/2}$ at low photon energies for the stoichiometric MnBi$_2$Te$_4$ film (1) (0.34\,eV), for the Bi$_2$Te$_3$ single crystal (0.31\,eV), and for the MnTe phase (1.41\,eV) (see Fig.\,\ref{Gaps}(a)). The direct gap values were estimated from the linear extrapolation of the $[\alpha(\omega)\omega]^2$ from the low to high photon energies for the stoichiometric MnBi$_2$Te$_4$ film (1) (1.20\,eV, 2.05\,eV), for the Bi$_2$Te$_3$ single crystal (1.12\,eV, 1.67\,eV), and for the MnTe phase (2.70\,eV) (see Fig.\,\ref{Gaps}(b)). We would like to note that the determined indirect optical gap value of $\sim$1.4\,eV for MnTe layer is in good agreement with the experimentally determined optical gap for MnTe single crystal \cite{Allen, Ferrer}. The estimated gap values are summarized in Table\,\ref{tab:table2}.

\begin{table} [h]
\caption{\label{tab:table2}
Parameters of the Gaussian oscillators (a) for the stoichiometric MnBi$_2$Te$_4$ film (1), (b) for the MnTe phase resulting from the EMA model, and (c) for the Bi$_2$Te$_3$ single crystal (see the text for details). The indirect and direct gap values are determined in Fig.\,\ref{Gaps}.}
\begin{ruledtabular}
\begin{tabular}{lccccc}
\,\,\,\,\,\,\,\,\,\,             & $A$  & $E_j$ & $Br$ & Indirect\,gap & Direct\,gap \\
                                 &      &(eV)   & (eV) &   (eV) &    (eV)   \\ \hline
(a) MnBi$_2$Te$_4$               &19.4  & 1.13  & 1.04 & 0.34 & 1.20 \\
                                 &14.3  & 2.1   & 1.8  & & 2.05 \\
                                 &1.2   & 3.5   & 1.2  & & \\
                                 &4.9   & 3.6   & 3.5  & & \\ 
                                 &3.0   & 7.2   & 2.7  & &\\ \hline
(b) MnTe                         & 5.1  & 2.5   & 0.9  &1.41 & 2.70\\
                                 & 9.4  & 3.0   & 1.7  & & \\
                                 & 5.9  & 4.1   & 2.5  & & \\
                                 & 4.0  & 7.1   & 2.3  & & \\ \hline
(c) Bi$_2$Te$_3$                 &150.0 & 0.17  & 3.4  &0.31 & 1.12\\
                                 &33.6  & 1.1   & 1.0 & & 1.67\\
                                 &3.6   & 1.5   & 0.5  & &  \\
                                 & 2.7  & 5.8   & 10.0 & & \\ 
\end{tabular}
\end{ruledtabular}
\end{table}  

\begin{figure*} [!ht] \centering
\includegraphics[width=13.5cm]{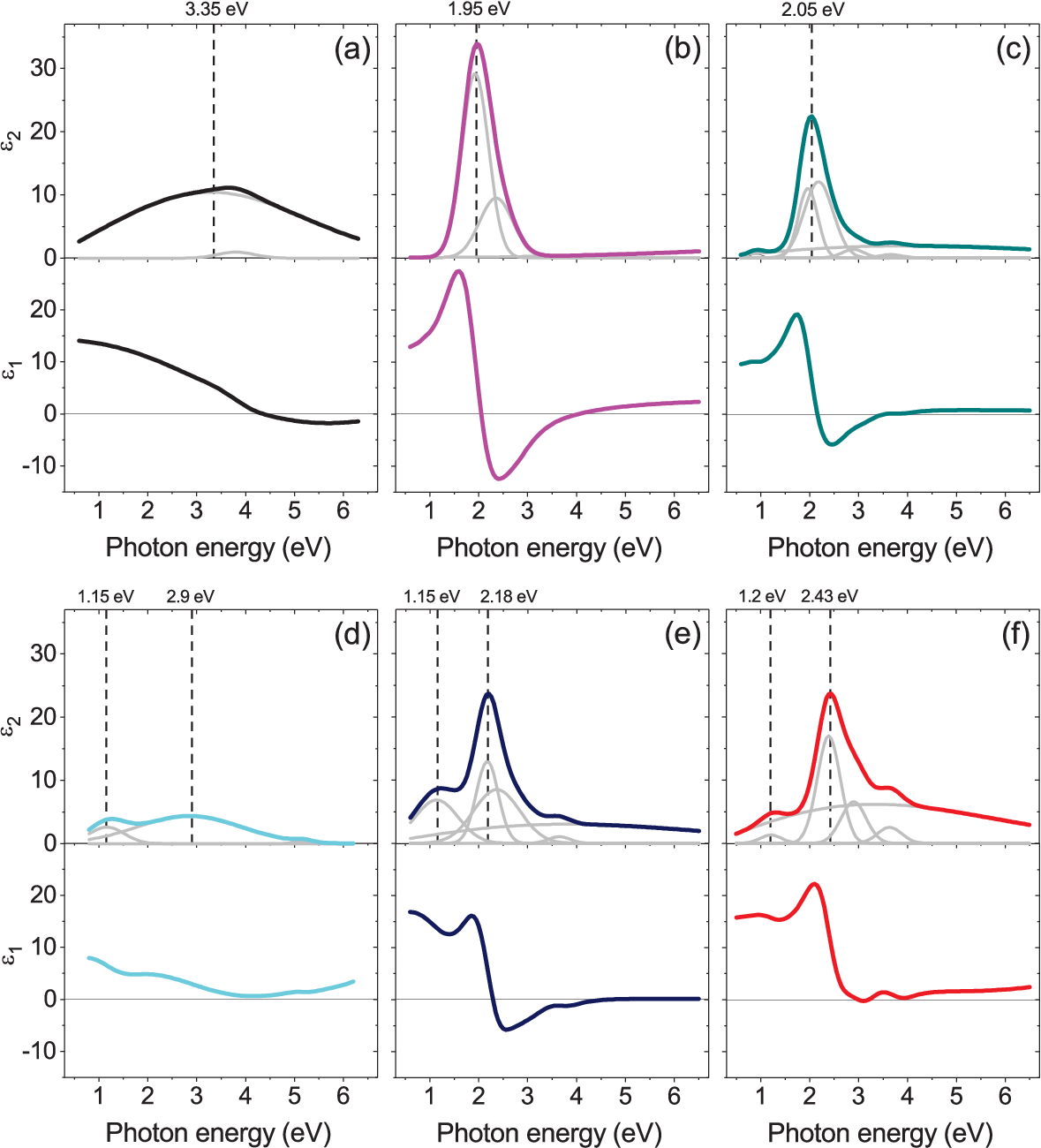}
\caption{(a-f) Imaginary (top panels) and real (bottom panels) parts of the complex dielectric function, $\varepsilon_2(\omega)$ and $\varepsilon_1(\omega)$, of the defect-generated phase (EMA 3rd constituent) determined from the EMA simulation for the film samples (2) MnBi$_{1.85}$Te$_{3.8}$/Si, (3) MnBi$_{1.6}$Te$_{2.9}$/Si, (4) Mn$_2$BiTe$_2$/Si, (5) Mn$_{1.3}$BiTe$_{2.5}$, (6) Mn$_2$BiTe$_4$, and (7) Mn$_2$Bi$_{1.2}$Te$_{4.2}$. The contribution from Gaussian optical bands is illustrated by solid gray curves (for more detail, see the text).
\vspace{-0.5cm}}
\label{CompDef}
\end{figure*}

\subsubsection{\label{sec:level2_1}Three-phase EMA model for the inhomogeneous MBE grown Mn-Bi-Te films}
The Bi$_2$Te$_3$ and MnTe phase constituents introduced in the two-phase EMA model for the stoichiometric MnBi$_2$Te$_4$ film (1) are presented in Fig.\,\ref{EMA_N164}(b,c), respectively. Then, using the three-phase Bruggeman EMA model complemented by the unknown (to be determined) defect-phase -- 3rd constituent, we model the MBE grown inhomogeneous Mn-Bi-Te films (2)-(7), and the Mn-Bi-Te thickness ($d_{\rm EMA}$) and effective thickness of the sr layer were fitted. The imaginary (top panels) and real (bottom panels) parts of the effective complex dielectric function, $\varepsilon_2(\omega)$ and $\varepsilon_1(\omega)$, respectively, of the defect-associated phase (3rd-phase constituent), determined from the EMA simulation of the ellipsometric angles, $\Psi(\omega)$ and $\Delta(\omega)$, measured for the film samples 
(2) MnBi$_{1.85}$Te$_{3.8}$, (3) MnBi$_{1.6}$Te$_{2.9}$, (4) Mn$_2$BiTe$_2$/Si, (5) Mn$_{1.3}$BiTe$_{2.5}$/Si, (6) Mn$_2$BiTe$_4$/Si, and (7) Mn$_2$Bi$_{1.2}$Te$_{4.2}$/Si are presented in Fig.\,\ref{CompDef}(a-f). 

From Fig.\,\ref{CompDef}(a) one can see that the defect-associated optical band is rather wide in the film (2) MnBi$_{1.85}$Te$_{3.8}$, which is indicative of a comparatively small defect concentration resulting in their non-uniform distribution. Here, the non-stoichiometric MnTe $\cdot$ Bi$_{(2-x)}$Te$_{3(1-x/2)}$ film composition corresponds to 
$x$\,$\sim$\,0.15, in agreement with the $\sim$\,8\% defect-phase contribution inside the $\sim$\,53\% Bi$_2$Te$_3$ phase (see Table\,\ref{tab:table1}). We suggest that vacancies [V$_{\rm Bi}$]$^{3-}$ and/or [V$_{\rm Te}$]$^{2+}$ in the structure of the MnTe and Bi$_2$Te$_3$ layers can be incorporated in these defects. Indeed, the energy levels associated with vacancies are generally located deep inside the valence band \cite{Kovaleva_JETP,Kovaleva_PhysB}, which can cause the localization of intrinsic charge carriers of $n$- and $p$-types, tuning the Fermi level $E_F$ to the middle of the gap. The film (5) Mn$_{1.3}$BiTe$_{2.5}$ reveals similar defect-related wide-band optical response (see Fig.\,\ref{CompDef}(d)). Earlier, it was shown that an excess of Ge in alloys Ge$_{1.2}$Bi$_2$Te$_4$ leads to the pronounced structural changes confirmed with a new system of XRD reflexes, which correspond to the nine-layer (NL) structure of Ge$_{1.5}$Bi$_{2.5}$Te$_5$, with the parameters of the hexagonal unit cell (space group $P\bar{3}m1$) $a$\,=\,4.305\,~$\AA$ and $c$\,=\,17.372\,$\AA$ (for more detail, see \cite{Karpinsky_1997}). The nominal composition in the film (5) Mn$_{1.3}$BiTe$_{2.5}$ corresponds to the slightly higher than 4/5 cation/anion ratio. An additional comprehensive XRD study (which is beyond the present optical study) is required to verify whether its structure corresponds to the NL structure. We can say that the non-stoichiometric film composition for the film MnBi$_{1.6}$Te$_{2.9}$ (3) is associated with the $\sim$13\% defect-phase inside the $\sim$42\% Bi$_2$Te$_3$ phase (see Table\,\ref{tab:table1}). As follows from the obtained EMA results, strong deviation from the stoichiometry for $x$\,$\geq$\,0.3 in MnTe$\cdot$Bi$_{(2-x)}$Te$_{3(1-x/2)}$ seemingly results in the formation of defects of different kinds in the film structure, which exhibits relatively narrow optical band(s) in the dielectric function spectra. Indeed, from Fig.\,\ref{CompDef}(b) one can see that the defect-phase constituent for the film (3) MnBi$_{1.6}$Te$_{2.9}$ displays the pronounced and relatively narrow optical band at $\sim$1.95\,eV. Moreover, from Fig.\,\ref{CompDef}(c,e,f) one can follow that the defect-associated optical response shifts to higher energies of $\sim$2.05, $\sim$2.18, and $\sim$2.43\,eV in the films (4) Mn$_2$BiTe$_2$, (6) Mn$_2$BiTe$_4$, and (7) Mn$_2$Bi$_{1.2}$Te$_{4.2}$, respectively. The discovered trend indicates that the defect-associated energy levels become more deeply located with increasing non-stoichiometry in the studied Mn-Bi-Te films.

\section{\label{sec:level1}IR STUDIES OF Mn-Bi-Te FILMS}
\begin{figure}
\includegraphics*[width=0.9\columnwidth]{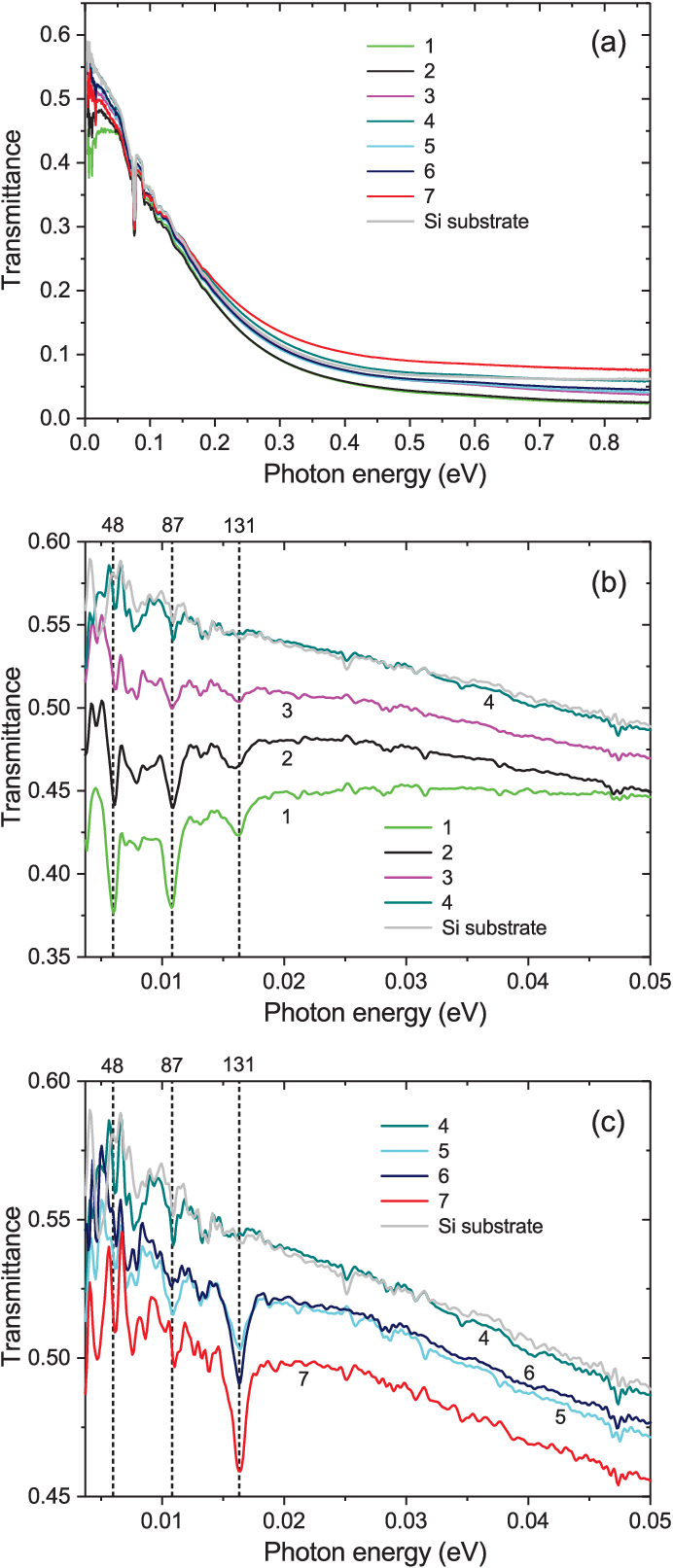}
\caption{(a) IR transmittance spectra for the SiO$_2$(2.5\,nm)/Si substrate and for the film samples (1) MnBi$_2$Te$_4$/Si, (2) MnBi$_{1.85}$Te$_{2.8}$/Si, (3) MnBi$_{1.7}$Te$_{2.55}$/Si, (4) Mn$_2$BiTe$_2$/Si, (5) Mn$_2$Bi$_2$Te$_5$/Si, (6) Mn$_2$Bi$_{1.5}$Te$_{4.2}$/Si, and Mn$_2$Bi$_{1.4}$Te$_4$/Si. (b,c) The transmittance spectra for the film samples (1)-(4) and (4)-(7) in the FIR, respectively.
\vspace{-0.5cm}}
\label{Phonons}
\end{figure}

The 30-7000\,cm$^{-1}$ (0.004-0.9\,eV) IR transmittance was measured for the investigated series of Mn-Bi-Te/Si film samples (with 2\,cm$^{-1}$ spectral resolution) using a VERTEX 80v FTIR spectrometer (see Fig.\,\ref{Phonons}(a)). In addition, the reference transmittance of the bare SiO$_2$(2.5\,nm)/Si substrate was measured. Figure\,\ref{Phonons}(b) displays more details in the FIR range for the film samples (1) MnBi$_2$Te$_4$, (2) MnBi$_{1.85}$Te$_{3.8}$, (3) MnBi$_{1.6}$Te$_{2.9}$, and (4) Mn$_2$BiTe$_2$. As one can see from the FIR transmittance spectra presented in Fig.\,\ref{Phonons}(b), the stoichiometric MnBi$_2$Te$_4$ film shows three IR-active phonons at 48, 87, and 131\,cm$^{-1}$, corresponding to the IR-active phonons showing up at 47, 84, and 133 cm$^{-1}$ in MnBi$_2$Te$_4$ single crystal \cite{Cristian}. Obviously, with decreasing Bi and Te contents, the IR phonon features at 48, 87, and 131 cm$^{-1}$ progressively decrease in the films MnBi$_{1.85}$Te$_{3.8}$ (2) and MnBi$_{1.6}$Te$_{2.9}$ (3), and, eventually, the IR phonon features disappear in the Mn$_2$BiTe$_2$ film (4). According to the obtained EMA results (see Table\,\ref{tab:table1}), the film (4) contains $\sim$46\% of the MnTe phase, while the volume fractions of the Bi$_2$Te$_3$ and the defect-associated phases are approximately equal. Then, for this phase constituents the film Mn$_2$BiTe$_2$ (4) seemingly becomes structurally disordered. As one can see from the FIR transmittance spectra presented in Fig.\,\ref{Phonons}(c), an increase of the MnTe content in the Mn-Bi-Te films (4)
 Mn$_2$BiTe$_2$, (5) Mn$_{1.3}$BiTe$_{2.5}$, (6) Mn$_2$BiTe$_4$\,=\,MnTe\,$+$\,MnTe$\cdot$BiTe$_2$, and (7) Mn$_2$Bi$_{1.2}$Te$_{4.2}$\,=\,MnTe\,$+$\,MnTe$\cdot$Bi$_{1.2}$Te$_{2.2}$ leads to the reinforcement of the phonon feature at 131\,cm$^{-1}$, which is seemingly associated here with the IR phonon observed at 131\,cm$^{-1}$ in MnTe single crystal \cite{Allen}. This trend is especially pronounced for the film (7) MnTe\,$+$\,MnTe$\cdot$Bi$_{1.2}$Te$_{2.2}$, where according to the EMA results (see Table\,\ref{tab:table1}) the MnTe content acquires $\sim$\,54\%. Meantime, whilst the phonons at 48 and 87\,cm$^{-1}$ cannot clearly be distinguished in the FIR spectra of the film samples (4) Mn$_2$BiTe$_2$, (6) Mn$_2$BiTe$_4$ and (7) Mn$_2$Bi$_{1.2}$Te$_{4.2}$, the IR phonon at 87\,cm$^{-1}$ is definitely present in the film (5) Mn$_{1.3}$BiTe$_{2.5}$. We think, thic gives an additional indication on the existence of the ordered NL phase in this film.  

In addition, from Fig.\,\ref{Phonons}(b) one can see that with decreasing photon energy the low-energy transmittance notably decreases for the stoichiometric film MnBi$_2$Te$_4$ (1) with respect to the Si substrate. The decreased transmittance can be ascribed to the Drude-type contribution(s). The low-energy response of MnBi$_2$Te$_4$ single crystal with the 
N\'{e}el temperature $T_N$\,$\simeq$\,25\,K has recently been investigated by IR spectroscopy \cite{Cristian}. From the quantitative Drude-Lorentz analysis of the IR optical conductivity $\sigma_1(\omega)$, two Drude terms describing the itinerant charge carriers response were identified, each characterized by a plasma frequency $\omega_{\rm{pl}}$ and a scattering rate $\gamma_D$ ($\omega_{{\rm pl}}$\,=\,6215\,cm$^{-1}$, $\gamma_D$\,=\,520\,cm$^{-1}$ and $\omega_{{\rm pl}}$\,=\,1870\,cm$^{-1}$, $\gamma_D$\,=\,150\,cm$^{-1}$). The total estimated charge carrier concentration is $n=n_1+n_2$\,=\,1.7$\times$10$^{20}$ cm$^{-3}$, where $n_1$\,$\simeq$\,0.517$\times$10$^{20}$\,cm$^{-3}$ and $n_2$\,$\simeq$\,1.183$\times$10$^{20}$\,cm$^{-3}$ \cite{Cristian}, which, in accordance with Eq.\,(\ref{eq1}), can be associated with the presence of the antisite Te$^{\circ}_{\rm Bi}$ and substitutional Bi$^{\circ}_{\rm Mn}$ donor-type defects, respectively. The simulated transmittance accounting for the absorption of the Drude charge carriers associated with the intrinsic Te$^{\circ}_{\rm Bi}$ and Bi$^{\circ}_{\rm Mn}$ $n$-type doping is in good agreement with the FIR transmittance for the stoichiometric film MnBi$_2$Te$_4$ (1) (for more detail, see Supplementary online information in \cite{Kovaleva_APL_2024}). The FIR transmittance apparently increases for the films MnBi$_{1.85}$Te$_{3.8}$ (2) and MnBi$_{1.6}$Te$_{2.9}$ (3) possessing the reduced Bi and Te stoichiometry. This effect can be explained by the decreased free charge carrier concentration associated with the intrinsic $n$-type doping. In addition, the measured FIR transmittance spectra for the Mn-Bi-Te films (5) Mn$_{1.3}$BiTe$_{2.5}$, (6) Mn$_2$BiTe$_4$\,=\,MnTe\,$+$\,MnTe$\cdot$MnBiTe$_2$, and (7) Mn$_2$Bi$_{1.2}$Te$_{4.2}$\,=\,MnTe\,$+$\,MnTe$\cdot$MnBi$_{1.2}$Te$_{2.1}$ with the increased MnTe content clearly do not exhibit a decreased FIR transmittance at $\omega \rightarrow 0$, which is indicative of the substantially reduced (or absent) contribution(s) from the free charge carriers.

\section{\label{sec:level1}SUMMARY}
By using 0.5-6.5\,eV spectroscopic ellipsometry, we investigated the complex dielectric function spectra, $\tilde\varepsilon(\omega)$=$\varepsilon_1(\omega)$+i$\varepsilon_2(\omega)$, for the series of the MBE grown Mn-Te-Bi film samples onto Si(111) substrates with the decreased Bi and Te contents and increased Mn content relative to the nominal stoichiometry in the AFM topological insulator (TI) MnBi$_2$Te$_4$. In addition, the 0.004-0.9\,eV IR transmittance spectra were measured for the studied film samples. The measured FIR transmittance spectra for the non-stoichiometric Mn-Bi-Te films show substantially reduced (or absent) contribution(s) from the free charge carriers, which supports the relevance of the localization effects. An effective medium approximation (EMA) was used to reproduce the complex dielectric function of the stoichiometric MnBi$_2$Te$_4$ film in terms of the two-phase -- Bi$_2$Te$_3$ and MnTe -- composite. The results obtained for the inhomogeneous Mn-Bi-Te films using the three-phase EMA model indicate that the defect-associated optical response systematically shifts to the higher photon energies from $\sim$1.95 to $\sim$2.43\,eV with decreasing Bi and Te contents and increasing Mn content, pointing out that the electrons become more deeply localized. The discovered trend indicates that the structural inhomogeneities in the non-stoichiometric Mn-Bi-Te films form holes in the Bi and Te layered structure, which can be associated with quantum anti-dots (QADs).  

Provided that the deficiency of Bi and Te ions is appreciable, the mosaics of QADs can be formed within the Bi and Te layers, which, as a result, will look like being perforated. The morphology profile corresponding to the formed QADs mosaics will naturally be reproduced along the $c$ axis within the SL structure. The formed potential of inhomogeneously distributed QADs will necessarily lead to the localization of itinerant charge carriers originating from the intrinsic doping. The energy levels of electrons localized in the QADs could be recognizable under the stimulating optical excitation. By analogy, it was demonstrated that the top sheet layer of bilayer CVD graphene treated with nitrogen plasma becomes perforated and exhibits green emission \cite{Kovaleva_2D_Materials}.   

The obtained results indicate that the structure of the non-stoichiometruc films is not continuous but represented by the regions of nearly stoichiometric MnBi$_2$Te$_4$ phase, where the $E_F$ is located within the bulk gap, which includes hollows or quantum anti-dots (QADs). The formed QADs mosaics formed within the SL building blocks along the $c$ axis in the non-stoichiometric Mn-Bi-Te films will break the continuity conditions necessary for quantum transport measurements in the topological Dirac surface states. Provided that the percolation conditions are satisfied for the TI regions determined by the QADs size quantization, the tunneling will allow the observation of the QAH effect at low temperatures. 

In fact, the films under study can be treated as a novel quantum object: a mosaic of small-scale regions exhibiting topological features. Our results can provide additional prospects for the search for topological materials and for fine tuning of their characteristics.\\

This work was partially supported in the framework of the State assignment of the ISSP RAS and LPI RAS. The spectroscopic ellipsometry measurements were performed at the LPI RAS Shared facility center. The FTIR transmittance and AFM studies were fulfilled in the ISSP RAS Shared facility center.\\

The authors have no conflicts to disclose.\\

The data that support the findings of this study are available
within this article.

\end{document}